\documentclass{PoS}

\usepackage{wrapfig}

\def\JSNS2{JSNS$^2$}

\title{Status and Prospects of the JSNS$^2$ Experiment}

\ShortTitle{JSNS$^2$ Status}

\author{\speaker{Carsten Rott}\thanks{on behalf of the JSNS$^2$ collaboration}\\
        Department of Physics, Sungkyunkwan University, Suwon 16419, Korea\\
        E-mail: \email{rott@skku.edu}}

\abstract{The JSNS$^2$ experiment aims to search for the existence of neutrino oscillations with $\Delta {\rm m}^2$ near 1 eV$^2$ at the J-PARC Materials and Life Science Experimental Facility~(MLF). A 3~GeV 1~MW proton beam incident on a mercury target produces an intense neutrino source from muon decay at rest ($\mu^{+} \rightarrow e^{+} + \bar{\nu}_{\mu} + \nu_{e}$). The oscillation to be searched for is $\bar{\nu}_{\mu}$ to $\bar{\nu}_{e}$, detected via the inverse beta decay~(IBD) reaction ($\bar{\nu}_{e} + p \rightarrow e^{+} + n$), which is then distinctively tagged by gammas from neutron capture of Gadolinium. The first of two detectors with 17~tons fiducial volume is currently under construction at a distance of 24~m from the mercury target. JSNS$^2$ is expected to provide the ultimate test of the LSND anomaly by replicating nearly identical conditions. The status of the experiment, which is expected to start taking data in Spring 2019, is discussed and its physics potential reviewed.}

\FullConference{The 39th International Conference on High Energy Physics (ICHEP2018)\\
		4-11 July, 2018\\
		Seoul, Korea}

\begin{document}

\section{Introduction}

The JSNS$^2$ - J-PARC Sterile Neutrino Search at J-PARC Spallation Neutron Source (E56) experiment~\cite{Ajimura:2017fld} is currently under construction and will allow to search for sterile neutrinos from appearance of active flavors with a new $\Delta m^2$ inconsistent with $\Delta m^2_{12}$ or $\Delta m^2_{23}$. The experiment is motivated by a series of anomalies observed in preceding experiments, including decay-at-rest~(DAR), decay-in-flight~(DIF), radioactive-source, and reactor experiments, that are summarized in
table~\ref{tab:LSND}. 
\begin{table}[h]
\begin{center}
	\begin{tabular}{|l|c|c|c|l|}
	\hline
	Experiment     & $\nu$-source     & Energy~$E_{\nu}$ / Distance~$L$  & Signal     & Significance \\ \hline \hline
	LSND~\cite{LSND}               & $\pi$ DAR & 40~MeV / 30~m & $\bar{\nu}_{\mu}\rightarrow\bar{\nu}_e$  & $3.8~\sigma$     \\ \hline
	MiniBooNE~\cite{MiniBooNE2018,MiniBooNE2013}       & $\pi$ DIF & 800~MeV / 600~m    & $\nu_{\mu}\rightarrow\nu_e$ + $\bar{\nu}_{\mu}\rightarrow\bar{\nu}_e$   &  $4.8~\sigma$         \\ \hline
	Gallium/SAGE~\cite{Giunti:2010zu} & e capture         & $<3$~MeV / 10~m    & $\nu_e\rightarrow\nu_x$   & $2.7~\sigma$  \\ \hline
	Reactors~\cite{Mention:2011rk}   &  $\beta$ decay           & 3~MeV / 10-100~m    & $\bar{\nu}_e\rightarrow\bar{\nu}_x$ & $3.0~\sigma$   \\ 
	\hline
	\end{tabular}
	\caption{Summary of observed experimental anomalies and their relative significances that hint at a possible large $\Delta m^2$.}
        \label{tab:LSND}
\end{center}
\end{table}

One of the longest standing and most significant neutrino anomalies was observed in data collected with the Liquid Scintillator Neutrino Detector~(LSND) at Los Alamos in 1998~\cite{LSND}. Based on an observed excess of $\bar{\nu_e}$ events by LSND follow-up experiments (such as MiniBooNE) were conducted, which found further anomalies. While, it remains unclear whether the excesses are due to oscillations, if interpreted as such both LSND and MiniBooNE indicate a flavor conversion of $\bar{\nu}_{\mu}$ to $\bar{\nu}_e$ at a probability of about $0.003$ with a $\Delta m^2$ of $\sim 1$~eV$^2$~\cite{Collin:2016aqd}. The process is however disfavored based on atmospheric neutrino oscillations~\cite{Aartsen:2017bap}.

JSNS$^2$ will perform a direct test of the long standing anomaly using the same neutrino source, target material and neutrino oscillation baseline length. Besides the main science goal, the discovery of sterile neutrinos or excluding their existence based on the evidence presented by previous experiments, JSNS$^2$ will be able to cover a large variety of scientific topics. Guaranteed results include the measurement of the neutrino cross section of $\sigma(^{12}C(\nu_{e},e^{-})^{12}N)$, which is critical for the understanding of core collapse supernovae ; precision measurement of monoenergetic 236~MeV neutrinos from KDAR ($K^+ \rightarrow \mu^+ \nu_\mu$; BR=63.5\%), which was recently observed by MiniBooNE~\cite{Aguilar-Arevalo:2018ylq} and is also relevant for dark matter searches~\cite{Rott:2015nma}; JSNS$^2$ also holds the potential to directly search for Pseudo-Dirac dark matter~\cite{Jordan:2018gcd}.

\section{The JSNS$^2$ Detector}

The JSNS$^2$ detector is a 50~ton liquid scintillator~(LS) detector using Linear Alkyl Benzene (LAB) as base solvent. Seventeen~tons of the LS is Gd-loaded and contain in an acrylic vessel of 3.2~m diameter and 2.5~m height to form the inner detector. This neutrino target is observed by 192~10-inch Hamamatsu R7081 PMTs ($5\times24$~wall PMTs and $2\times 36$ top/bottom PMTs) attached to a holding structure on the inner part of a stainless steel tank with 4.6~m height and 3.5~m diameter containing LS. 
      A LS filled veto layer between the tank wall and PMT holding structure of 25~cm to 45~cm thickness is viewed by a total of 48~10-inch PMTs with 12~PMTs each on the top and bottom layers and $2\times 12$~PMTs in the side region.
CAEN V1730 Flash ADCs are employed as wave-form digitizers for the PMT signals to provide 14-bit 500-MHz sampling. The detector is expected to provide an energy resolution better than 2.5\% at 50~MeV.

\begin{figure}[t]
        \centering
        \includegraphics[width=0.355\textwidth]{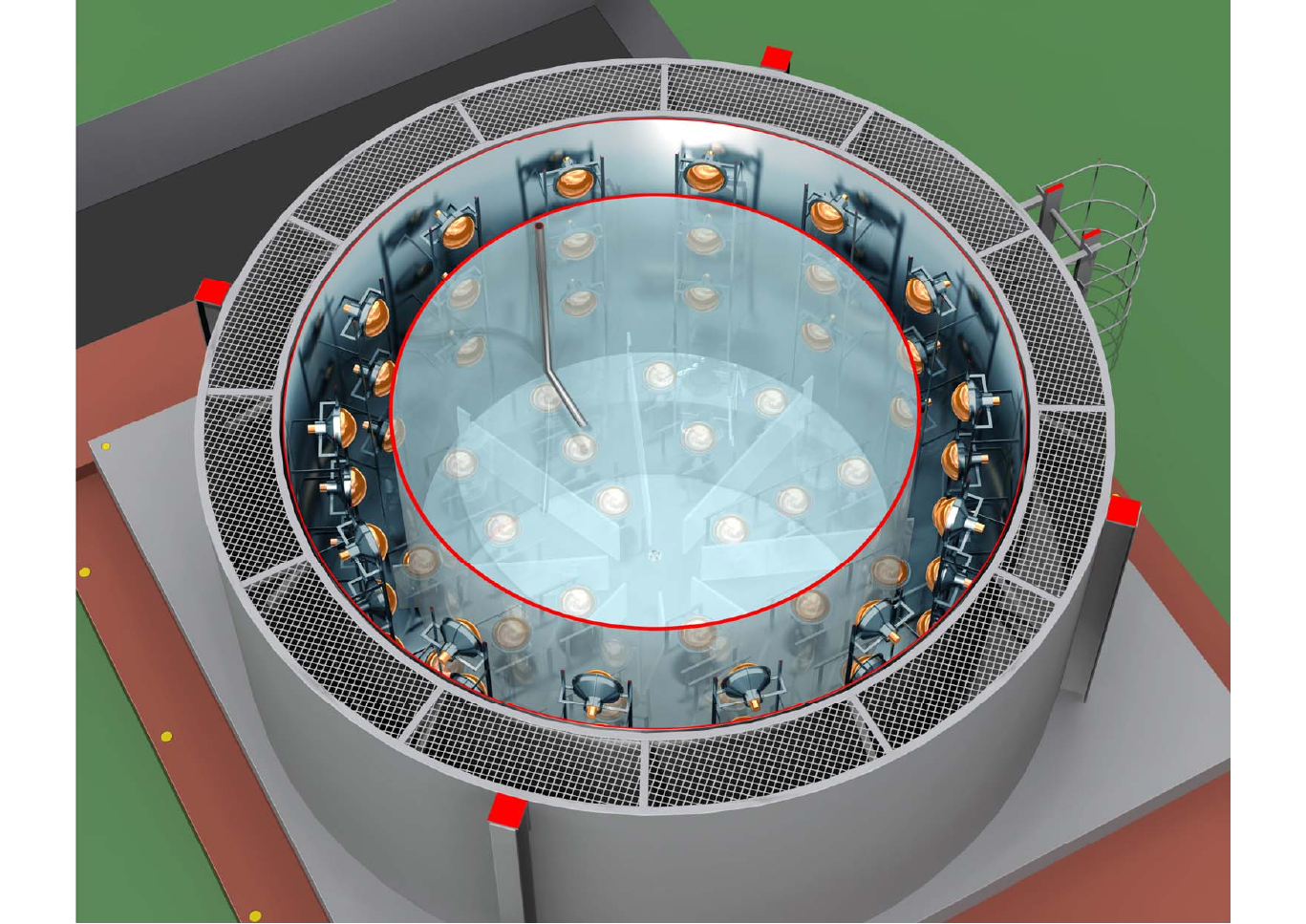}
        \includegraphics[width=0.385\textwidth]{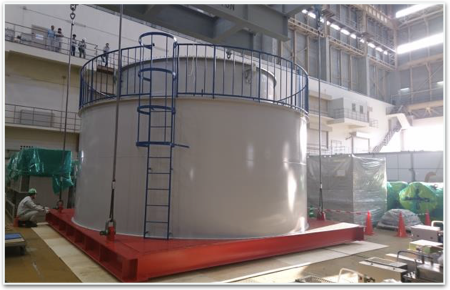}
        \includegraphics[width=0.185\textwidth]{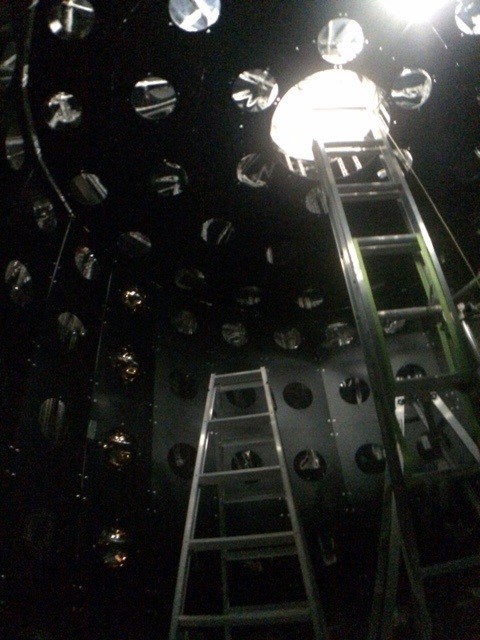}
        \includegraphics[width=0.77\textwidth]{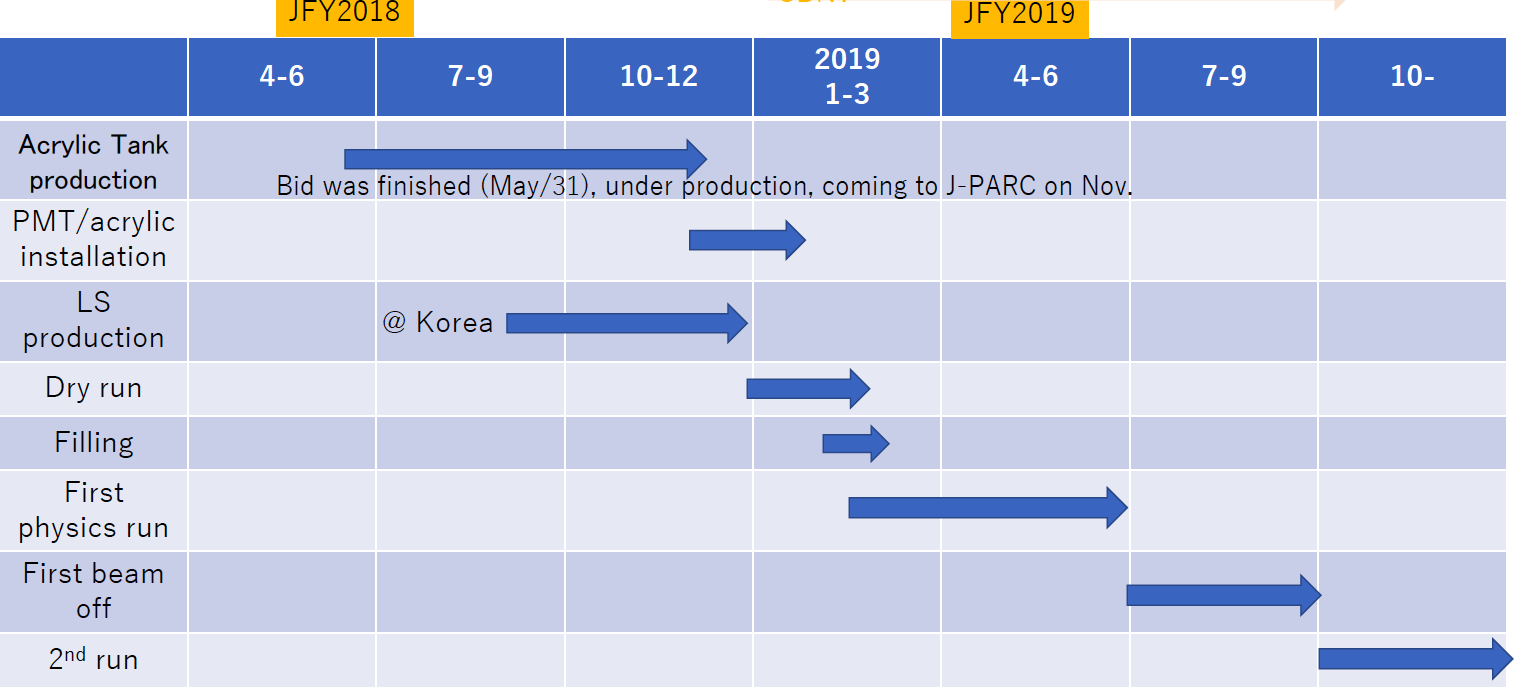}
        \caption{Top left: JSNS$^{2}$ detector schematics; Top middle: JSNS$^{2}$ detector in the HENDEL assembly hall (March 2018); Top right: Inside tank (September 2018) ; Bottom: The JSNS$^{2}$ construction schedule.
        }
        \label{JSNS2_status}
\end{figure}

\section{Sensitivity}

\begin{figure}[htb]
        \centering
       \includegraphics[width=0.92\textwidth]{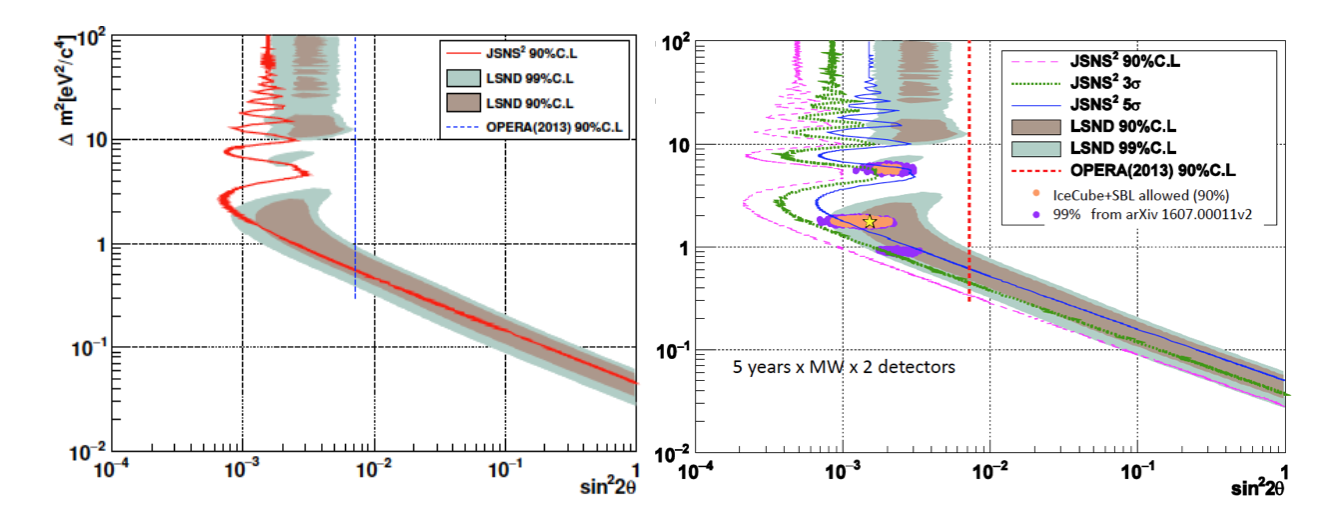}
        \caption{Sensitivity and discovery prospects of the JSNS$^2$ experiment.
        }
        \label{JSNS2_sensitivity}
\end{figure}

JSNS$^2$ was designed to discover sterile neutrinos with 5~years of data or  exclude the LSND anomaly at 90\% C.L. with 3~years of data. The sensitivity and discovery potential is shown in Figure~\ref{JSNS2_sensitivity}. JSNS$^2$ substantially improvements over LSND in numerous ways: (1) MLF provides a high intensity short pulsed beam significantly reducing the coincidental backgrounds; (2) Gd-doped LS allows to tag IBD events; 
(3) Contrarily to LSND the JSNS$^2$ detector is located above the beam dump reducing the possibility of beam related backgrounds that has been suggested as a potential background to the LSND measurement.

\section{Conclusions}

Global neutrino data is inconclusive on the existence of sterile neutrinos. During the past two decades, sterile neutrino searches were driven by the LSND anomaly and results obtained in follow up experiments. The JSNS$^2$ experiment is a direct and ultimate test of this long standing anomaly, by using the same neutrino source, target material and baseline length. JSNS$^2$ has significant experimental improvements over LSND, including a factor 100~S/N increase and reduced systematic uncertainties.
JSNS$^2$ will start data taking in 2019 and will be able to discover (exclude) a sterile neutrino signal as indicated by LSND with 5 years (3years) of data. The JSNS$^2$ TDR~\cite{Ajimura:2017fld} describes the experiment and science potential in detail.

\section{Acknowledgements}
I would like to thank my collaborators for comments and suggestions on these proceedings. CR acknowledges support by the National Research Foundation of Korea (NRF) funded by the Korea government (MSIP) Basic Science Research Program (NRF-2017K1A3A7A09015973).

\end{document}